\documentclass[12pt,draftclsnofoot,onecolumn,journal]{IEEEtran}
\usepackage{lineno}
\usepackage{amssymb,amsmath,stfloats,cite}
\usepackage[normalem]{ulem}
\usepackage{graphicx,subfigure,multirow}
\usepackage{setspace}
\usepackage{dsfont}
\usepackage{soul}
\usepackage{epsfig}
\usepackage{epstopdf}
\newtheorem{lemma}{\textbf{Lemma}}
\newtheorem{Prop}{\textbf{Proposition}}
\newtheorem{Def}{\textbf{Definition}}
\newtheorem{remark}{\textbf{Remark}}

\allowdisplaybreaks[4]
\usepackage{times,amsmath,color,amssymb,epsfig,cite,subfigure,algorithm,algorithmic}

\begin{document}
\title{Energy-Efficient Hybrid Precoding for\\ Massive MIMO mmWave Systems\\ With a Fully-Adaptive-Connected Structure}

\author{{Xuan Xue, \IEEEmembership{Member, IEEE}, Yongchao Wang, \IEEEmembership{Member, IEEE},
 \\
 Long Yang, \IEEEmembership{Member, IEEE}, \\ Jia Shi, \IEEEmembership{Member, IEEE}, and Zan Li, \IEEEmembership{Senior Member, IEEE}}

\renewcommand{\baselinestretch}{1.5}

\thanks{This paper was partially presented at the IEEE ICC 2019\cite{XuanICC2019}.}
}

\maketitle

\begin{abstract}
This paper investigates the hybrid precoding design in millimeter-wave (mmWave) systems with a fully-adaptive-connected precoding structure, where a switch-controlled connection is deployed between every antenna and every radio frequency (RF) chain. To maximally enhance the energy efficiency (EE) of hybrid precoding under this structure, the joint optimization of switch-controlled connections and the hybrid precoders is formulated as a large-scale mixed-integer non-convex problem with high-dimensional power constraints. To efficiently solve such a challenging problem, we first decouple it into a continuous hybrid precoding (CHP) subproblem. Then, with the hybrid precoder obtained from the CHP subproblem, the original problem can be equivalently reformulated as  a discrete connection-state (DCS) problem with only 0-1 integer variables.
For the CHP subproblem, we propose an alternating hybrid precoding (AHP) algorithm. Then, with the hybrid precoder provided by the AHP algorithm, we develop a matching assisted fully-adaptive hybrid precoding (MA-FAHP) algorithm to solve the DCS problem.
It is theoretically shown that the proposed MA-FAHP algorithm always converges to a stable solution with the polynomial complexity. Finally, simulation results demonstrate that the superior performance of the proposed MA-FAHP algorithm in terms of EE and beampattern.
\end{abstract}


\section{Introduction}\label{Sec:introduction}
By utilizing plentiful available spectral resources between 30GHz and 300GHz, millimeter wave (mmWave) communication has shown its great potentials for providing ultra-high data-rate transmission as well as massive device connectivity in the upcoming 5G systems \cite{mmWaveCOMag2014Feb,mmwave_introduction,mm-wave_it_works}. However, mmWave \textcolor{black}{links} are very sensitive to path loss attenuation and blockage, \textcolor{black}{thereby naturally being} limited in short-range communications \cite{HeathJSTSP2016,HanComMag2015}. On the other hand, thanks to the very short wavelength of mmWave signal, large-scale antenna arrays can be \textcolor{black}{embedded into small devices, thereby naturally forming \emph{massive MIMO mmWave systems}} \cite{mm-wave_potential}. Moreover, when proper precoding design is employed in such systems, both the capacity and reliability of mmWave transmission can be enhanced significantly \cite{AlkhateebComMag2014}.

Since large-scale antenna arrays are deployed at massive MIMO mmWave systems, the conventional \emph{full-digital} precoding architecture, in which each antenna connects to a dedicated radio frequency (RF) chain, may not be practical due to the excessive power consumption and unacceptable hardware complexity\cite{YooJSAC2006,RanganProceed2014,Hong2014ComMag}. To cope with this problem, \emph{hybrid} precoding architecture was introduced into massive MIMO mmWave systems in some recent studies \cite{sparse_precoding,Alter_Hybrid,Alkhateeb2015TWC,Liang2014WCL,Ni2016TCOM,chen2015WCL,Du2018TWC,Gao2016JSAC,Dai2015ICC,Jin2018TWC,Park2017TWC,Jing2018ComL,Yu2018JSTSP,Payami2016TWC}.
Different from the conventional full-digital precoding architecture, the \emph{hybrid precoding} architecture divides the overall precoding process into digital precoding and analog precoding, where the former uses much less RF chains than the full-digital precoding and the latter is realized by low-cost phase-shifters.
Compared with the conventional full-digital precoder, the hybrid precoder can achieve similar performance at much less hardware cost and power consumption.

According to the connecting method between the antennas and RF chains, the structure of hybrid precoding can be classified into  {the following three types}:
\begin{itemize}
\item \textcolor{black}{ {\textbf{Full-connected structure}, in which each RF chain is fixedly connected to all antennas, can achieve high spectral efficiency (SE) with properly designed precoding schemes.} The authors in \cite{sparse_precoding,Alter_Hybrid,chen2015WCL,Jin2018TWC} investigated the hybrid precoding in single-user massive MIMO mmWave systems.  {By exploting the sparse characteristics of mmWave channel, a low-complexity orthogonal matching persuit (OMP) algorithm was proposed in} \cite{sparse_precoding}.
     {Further, it was demonstrated in} \cite{Alter_Hybrid,chen2015WCL,Jin2018TWC}  {that the hybrid precoder can achieve similar SE performance to the optimal full-digital precoder.} When multiple users are simultaneously served in massive MIMO mmWave systems, several multiuser full-connected hybrid precoding schemes were proposed in \cite{Alkhateeb2015TWC,Ni2016TCOM,Liang2014WCL} to maximize the overall SE of all users.}
\item \textcolor{black}{ {\textbf{Sub-connected structure}, in which each RF chain is
fixedly connected to a certain portion of antennas, can consume less power than the full-connected structure at the expense of some spectral efficiency loss.} The works in \cite{Alter_Hybrid,Du2018TWC,Gao2016JSAC,Dai2015ICC} investigated the optimal hybrid precoding design targeting at SE maximization and evaluated the SE gap between sub-connected and full-connected structures.}

\item {\textbf{Adaptive-connected structure}, which employs the dynamically changing connections between RF chains and antennas, owns better flexibility for realizing hybrid precoding than the full-connected and sub-connected structures. Considering that the RF chains and the antennas are partitioned into a same number of groups, authors in \cite{Park2017TWC,Jing2018ComL,Yu2018JSTSP} designed hybrid precoders with the \emph{partially-adaptive-connected structure},  {where each group of RF chains is dynamically connected with a certain group of antennas. It was shown in} \cite{Park2017TWC,Jing2018ComL,Yu2018JSTSP}  {that the partially-adaptive-connected structure achieves a good tradeoff between SE and power consumption.}     On the other hand, the work in \cite{Payami2016TWC}
     {proposed to implement the hybrid precoder with} \emph{a fully-adaptive-connected structure},  {where the switch-controlled connections are deployed between the RF chains and the antennas.}  Since the on-off states of switch-controlled connections can be modified separately,  {the fully-adaptive-connected structure can smoothly switch among the full-connected structure and all possible sub-connected structures.}}
\end{itemize}
All aforementioned existing papers on hybrid precoding only investigate the SE maximization. However, in wireless communication systems, energy efficiency (EE), besides SE, is  another important performance metric, especially in  green communications for 5G/beyond \cite{Yazdan2017MicMag}.
Although some existing studies have proposed several hybrid precoding strategies to improve EE of massive MIMO mmWave systems\cite{He2017TSP,He2016access,Zi2016JSAC}, all of them are limited in the full/sub-connected structures with fixed connections.
Since the adaptive-connected hybrid precoding owns flexible hardware structure,
it could be more likely to achieve higher EE if RF chain-antenna connections and hybrid precoder are optimized jointly. However, the research on such a joint optimization for EE maximization of massive MIMO mmWave systems is still limited, which motivates us to investigate this issue.

 {In this paper, we focus on the joint optimization of RF chain-antenna connections and hybrid precoder under a fully-adaptive-connected structure, in order to maximally improve the EE of massive MIMO mmWave systems.} The main contributions are summarized as follows.
\begin{itemize}

\item \textcolor{black}{ {To maximize the EE of hybrid precoding with a fully-adaptive-connected structure, the joint optimization of switch-controlled connections and hybrid precoder is formulated as a large-scale high-dimensional mixed-integer non-convex problem.
    To efficiently solve such a challenging problem, we first decouple the original problem by assuming given switch-controlled connections and thereby obtain a continuous hybrid precoding (CHP) subproblem. Then, by regarding the optimal solution of CHP subproblem as a function of the given switch controlled connections, the original problem can be equivalently reformulated as a discrete connection-state (DCS) problem with only 0-1 integer variables.}}

\item For the CHP subproblem, we develop an alternating hybrid precoding (AHP) algorithm to optimize the hybrid precoder under given switch-controlled connections. Then, based on the AHP algorithm,  a matching assisted fully-adaptive hybrid precoding (MA-FAHP) algorithm is proposed to efficiently solve the DCS problem.

\item The convergence and complexity of the proposed MA-FAHP algorithm are theoretically analyzed for the proposed MA-FAHP algorithm. It is shown that the proposed MA-FAHP algorithm always converges to a stable solution with polynomial complexity. Simulation results demonstrate that the proposed MA-FAHP algorithm can achieve superior performance in terms of EE and beampattern.
\end{itemize}

The rest of this paper is organized as follows. Section \ref{Sec:system} describes the system model and the power consumption of the considered system. Section \ref{sec:formulation}  {formulates the joint optimization problem of hybrid precoders and switch-controlled connections and equivalently reformulates this problem to a DCS problem.}
Section \ref{sec:algorithm} proposes the MA-FAHP algorithm to solve the DCS problem. Section \ref{Sec:perforAnaly} theoretically analyzes the the convergence and complexity of proposed MA-FAHP algorithm. The simulation results are provided in Section \ref{Sec:simulation}, followed by Section \ref{sec:conclusion} that concludes this paper.

\emph{Notations}: Lowercase and uppercase boldface letters denote vectors and matrices, respectively; $(\cdot)^T$, $(\cdot)^H$, ${\rm Tr}(\cdot)$, ${\rm{vec}}({\cdot})$, and $\mathbb{E}[\cdot]$ symbolize the transpose, conjugate transpose, trace, vectorization and expectation operations, respectively; $||\mathbf{a}||_2$ denotes the 2-norm of vector $\mathbf{a}$; $||\mathbf{A}||_\mathbf{F}$ denote the Frobenius norm of matrix $\mathbf{A}$;   $\otimes$ and $\circ$  denote the Kronecker product and Hadamard product, respectively; ${\bf I}_{m}$ represents the $m\times m$ identity matrix; \textcolor{black}{$\{a\}^+$ symbolizes operation $\max\{0,a\}$}.

\section{System Model}\label{Sec:system}
This section first describes the system model for the fully-adaptive-connected hybrid precoding structure and then analyzes the power consumption of this structure.

\subsection{System Description}

\begin{figure}
\centering
\includegraphics[width=0.45\textwidth]{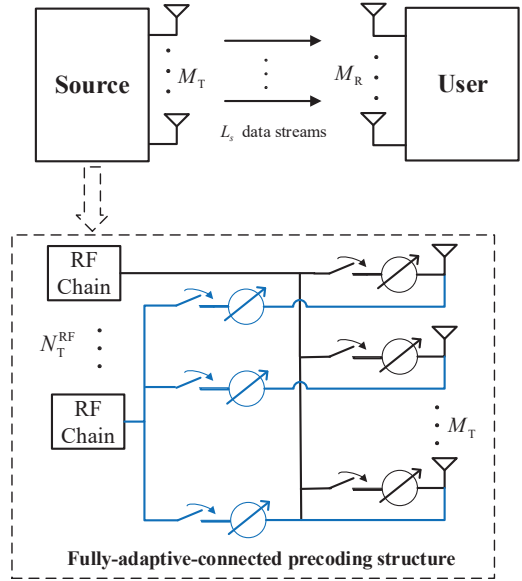}
\caption{The proposed fully-adaptive-connected precoding structure.}
\label{Fig:system}
\end{figure}
As shown in Fig. \ref{Fig:system}, a source \textcolor{black}{equipped} with $M_{\rm T}$ transmit antennas and $N_{\rm T}^{\rm RF}$ RF chains, intends to send a number $L_s$ of data streams to a user \textcolor{black}{equipped} with $M_{\rm R}$ receive antennas.
The hybrid precoding at the source is realized by a fully-adaptive-connected structure, where a switch-controlled connection is deployed between every RF chain and every antenna. It is worth to point out that, by modifying the on-off states of switch-controlled connections, this fully-adaptive-connected structure can realize a full-connected structure or any possible sub-connected structure. More precisely, if all connections switch to ``on'' state, \textcolor{black}{it} will work as a full-connected structure, otherwise, if some connections switch to ``off'' state, \textcolor{black}{it} will work as a certain sub-connected structure.

In each transmission, the data streams are precoded in this fully-adaptive-connected structure,
where the states of switch controlled connections are denoted by a 0-1 connection-state matrix $\bf{D}$. More precisely, defining the set of antennas as ${S}_{A}\triangleq\{1,\cdots,M_{\rm T}\}$ and the set of RF chains as ${S}_{RF}\triangleq\{1,\cdots,N_{\rm T}^{\rm RF}\}$, we denote ${\bf D}(i,j)=1$ or $0$ ($i\in{S}_{A}$ and $j\in{S}_{RF}$)  to represent that the switch-controlled connection between the $i$th antenna and  the $j$th RF chain is switched to ``on'' state or ``off'' state, respectively.
Further, denoting the digital precoder as ${\bf W}\in \mathcal{C}^{N_{\rm T}^{\rm{RF}}\times L_s}$ and the analog precoder as ${\bf F} \in \mathcal{C}^{M_{\rm T}\times N_{\rm T}^{\rm{RF}}}$, the transmit signal at the source can be expressed as
\begin{equation}\label{Eqn:transmit_signal}
{\bf x}=({\bf F} \circ{\bf D}){\bf W}\mathbf{s},
\end{equation}
where ${\bf s}\in \mathcal{C}^{L_s\times 1}$ is the transmitted symbol with satisfying $\mathbb{E}[{\bf ss}^H]=\frac{1}{L_s}{\bf I}_{L_s}$.
Considering that the channel matrix between the source and the user is ${\bf H}\in \mathcal{C}^{M_{\rm R}\times M_{\rm T}}$, the received signal at the user can be expressed as
\begin{equation}\label{Eqn:receive_signal}
{\bf y}={\bf H}{\bf x}+{\bf n}={\bf H}({\bf F} \circ{\bf D}){\bf W}\mathbf{s}+{\bf n},
\end{equation}
where ${\bf n}\in \mathcal{C}^{M_{\rm R}\times 1}$ represents the vector of zero-mean additive white Gaussian noise (AWGN) at the user. Without loss of generality, we assume the AWGN at each receive antenna has the same variance $\sigma^2$, i.e., ${\bf n}\sim \mathcal{CN}({\bf{0}}, \sigma^2{\bf I})$. Therefore, based on the received signal shown in \eqref{Eqn:receive_signal}, the achievable spectral efficiency can be expressed as
\begin{align}\label{Eqn:BSrate}
&\mathcal{R}({\bf W}, {\bf F}, {\bf D})=\\
&{\rm log}_2\Bigg[{\rm det}\Bigg({\bf I}
+\frac{{\bf H}({\bf F} \circ{\bf D}){\bf W}{\bf W}^H({\bf F} \circ{\bf D})^H{\bf H}^H}{L_s\sigma^2}\Bigg)\Bigg].\nonumber
\end{align}
\subsection{Power Consumption}
For the hybrid precoding with the fully-adaptive-connected structure, the overall power consumption includes both the \emph{transmit} power consumption and \emph{circuit} power consumption.
According to the transmit signal shown in \eqref{Eqn:transmit_signal}, the transmit power consumption can be obtained as
\begin{equation}\label{Eqn:tra_P}
P_t({\bf W}, {\bf F}, {\bf D})=\mathbb{E}(||{\bf x}||_2^2)
=\frac{1}{L_s}||({\bf F} \circ{\bf D}){\bf W}||_{\rm F}^2.
\end{equation}

On the other hand, the circuit power is consumed by four types of components: a) the \emph{working} RF chains, b) the \emph{working} phase shifters, \textcolor{black}{c) the switches,} d) the other circuit components whose power consumption does not change with hybrid precoding structure, including baseband processor, mixer, filter, etc. Without loss of generality, we assume that the circuit power consumption of the fourth-type components is fixed to $P_o$. Then, the circuit power consumption of the working RF chains, the working phase shifters and the switches can be analyzed as follows.

\emph{Circuit power consumption of working RF chains}: In the considered fully-adaptive-connected structure, the $j$th RF chain is working only when it is connected to at least one antenna, i.e., $\max_{i=1,...,M_{\rm T}}\{{\bf D}(i,j)\}=1$ holds. Thus, the number of working RF chains is given by
\begin{equation}
N_{\rm work}^{\rm RF}=\sum_{j=1}^{N_\mathrm{T}^{\rm RF}}\left[\max_{i=1,...,M_{\rm T}}{\bf D}(i,j)\right].
\end{equation}
Assuming each working RF chain consumes the same circuit power $P_{\rm RF}$, the power consumption of all working RF chains is obtained as $N_{\rm work}^{\rm RF}P_{\rm RF}=\left(\sum_{j=1}^{N_{\rm T}^{\rm RF}}\max_{i=1,...,M_{\rm T}}\{{\bf D}(i,j)\}\right)P_{\rm RF}$.

\emph{Circuit power consumption of working phase shifter}: For the switch-controlled connection between the $i$th antenna and the $j$th RF chain, its phase shifter is working only when $\mathbf{D}(i,j)=1$ holds. Therefore, the number of working phase shifters $N_{\rm work}^{\rm PS}$ is equal to the number of ``1'' elements in matrix $\mathbf{D}$, indicating
\begin{equation}
N_{\rm work}^{\rm PS}=\sum_{j=1}^{N_{\rm T}^{\rm RF}}\sum_{i=1}^{M_{\rm T}}\mathbf{D}(i,j)=||\mathbf{D}||_{\rm F}^2.
\end{equation}
 Assuming that each working phase shifter consumes the same power $P_{\rm PS}$, the power consumption of working phase shifters is $N_{\rm work}^{\rm PS}P_{\rm PS}=\|{\bf D}\|^2_{\rm F}\cdot P_{\rm PS}$.

\emph{Circuit power consumption of switches}: Since the source employs $M_{\rm T}$ antennas and $N_{\rm T}^{\rm RF}$ RF chains, the total number of switch-controlled connections is $M_{\rm T}N_{\rm T}^{\rm{RF}}$. Assuming that the power consumed by controlling every switch is $P_{\rm SW}$, the total power consumption of controlling all these switches is $M_{\rm T}N_{\rm T}^{\rm{RF}} P_{\rm SW}$.

By summarizing the above results, the total circuit power consumption can be obtained as
\begin{align}\label{Eqn:circuitpowerBS}
P_c({\bf D})
=&\left(\sum_{j=1}^{N_\mathrm{T}^{\rm RF}}\max_{i=1,...,M_{\rm T}}\{{\bf D}(i,j)\}\right)P_{\rm RF}+||{\bf D}||_{\rm F}^2P_{\rm PS}+M_{\rm T}N_{\rm T}^{\rm{RF}} P_{\rm SW}+P_{o}.
\end{align}

Finally, combining (\ref{Eqn:tra_P}) and (\ref{Eqn:circuitpowerBS}), the overall power consumption of the source is derived as
\begin{align}\label{Eqn:powerconsumptionBS_final}
P({\bf W}, {\bf F}, {\bf D})
=&\frac{||({\bf F} \circ{\bf D}){\bf W}||_{\rm F}^2}{{L_s}}+\left(\sum_{j=1}^{N_\mathrm{T}^{\rm RF}}\max_{i=1,...,M_{\rm T}}\{{\bf D}(i,j)\}\right) P_{\rm RF}
\\
&+||{\bf D}||_{\rm F}^2P_{\rm PS}+M_{\rm T}N_{\rm T}^{\rm{RF}} P_{\rm SW}+P_{o}.
\nonumber
\end{align}

\section{Problem Formulation}
\label{sec:formulation}

This section first formulates the joint optimization problem of hybrid precoders and switch-controlled connections,  {aiming at the maximization of  EE. Then, we show that the original joint optimization problem can be equivalently reformulated as a DCS problem with only 0-1 integer variables.}

\subsection{Problem Formulation for EE Maximization}
According to \eqref{Eqn:BSrate} and \eqref{Eqn:powerconsumptionBS_final}, the EE of considered system can be expressed as\cite{NgoTCOM2013}
\begin{IEEEeqnarray}{l}\label{Eqn:JSEE}
{\rm EE}({\bf W}, {\bf F}, {\bf D}) \triangleq \frac{\mathcal{R}({\bf W}, {\bf F}, {\bf D})}{P({\bf W}, {\bf F}, {\bf D})}.
\end{IEEEeqnarray}
\textcolor{black}{In that case}, the joint optimization problem of hybrid precoders and switch-controlled connections can be formulated as
\begin{subequations}\label{Eqn:problemJSEE_maxpower}
\begin{align}
\max_{{\bf W}, {\bf F}, {\bf D}}\hspace{0.5cm} & {\rm EE}({\bf W}, {\bf F}, {\bf D}), \label{Eqn:obj_ori}\\
\hspace{0.2cm} {\rm s.t.} \hspace{0.5cm} &||({\bf F} \circ{\bf D}){\bf W}||_{\rm F}^2\leq P_{\rm max}, \label{Eqn:power_cons}\\
&{\bf D}(i,j)\in\{0,1\},\hspace{5mm} \forall i\in{S}_{A}, \forall j\in{S}_{RF},\label{Eqn:switch_cons}
\\
&\hspace{0cm} |{\bf F}(i,j)|=\begin{cases} \frac{1}{\sqrt{M_{\rm T}}},&{\text{if }}  {\bf D}(i,j)=1,\\ 0,&{\text{if }} {\bf D}(i,j)=0,\end{cases}
\nonumber
\\
&\hspace{5mm} \forall i\in{S}_{A}, \forall j\in{S}_{RF},\label{Eqn:analog_cons}\\
&\sum_{j=1}^{N_{\rm T}^{\rm RF}}{\bf D}(i,j)\leq p_i, \forall i\in{S}_{A},\label{Eqn:validRFAnta_cons}\\
&\sum_{i=1}^{M_{\rm T}}{\bf D}(i,j)\leq q_j, \forall j\in{S}_{RF},\label{Eqn:validantanna_RF_cons}
\end{align}
\end{subequations}
\newline
where  constraints \eqref{Eqn:power_cons}$\sim$\eqref{Eqn:validantanna_RF_cons} are explained as follows: \eqref{Eqn:power_cons} is the transmit power constraint with $P_{\rm max}$ being the maximal transmit power; \eqref{Eqn:switch_cons} is the 0-1 constraint of the ``on-off'' state of each switch-controlled connection; \eqref{Eqn:analog_cons} represents the constant-modulus constraints for working elements of the analog precoder;
\eqref{Eqn:validRFAnta_cons} and \eqref{Eqn:validantanna_RF_cons} represent the practical implementation complexity constraints. In specific, \eqref{Eqn:validRFAnta_cons} means that  the $i$th antenna can be connected with up to $p_i(\leq N_{\rm T}^{\rm RF})$ RF chains, while \eqref{Eqn:validantanna_RF_cons} means that the $j$th RF chain can be connected with $q_j(\leq M_{\rm T})$ antennas at most.

\subsection{Problem Reformulation}
From problem \eqref{Eqn:problemJSEE_maxpower}, we have the following observations:
1) The three variables $\{{\bf W}, {\bf F}, {\bf D}\}$ are coupled in both fractional objective function \eqref{Eqn:obj_ori} and  power constraint \eqref{Eqn:power_cons}; 2) The analog precoder should satisfy the non-convex constant-modulus constraints \eqref{Eqn:analog_cons}; 3) The hybrid precoding ${\bf W}$ and ${\bf F}$ are continuous, while ${\bf D}$ is discrete variable with satisfying the constraints \eqref{Eqn:switch_cons}, \eqref{Eqn:validRFAnta_cons} and \eqref{Eqn:validantanna_RF_cons}.Therefore, problem \eqref{Eqn:problemJSEE_maxpower} is a large-scale high-dimensional mixed-integer non-convex problem with both continuous and discrete variables, thereby it is extremely challenging to efficiently solve such a problem.
To facilitate the mathematical tractability, we will equivalently reformulate problem \eqref{Eqn:problemJSEE_maxpower} by the following procedure.

When connection-state matrix ${\bf D}$ is given, problem \eqref{Eqn:problemJSEE_maxpower} can be \textcolor{black}{decoupled} as a CHP subproblem with respect to only continuous variables ${\bf W}$ and ${\bf F}$, which is given by
\begin{align}\label{Eqn:problemCHP1}
\max_{{\bf W}, {\bf F}}\hspace{0.5cm} & {\rm EE}({\bf W}, {\bf F})\nonumber\\
\hspace{0.2cm} {\rm s.t.} \hspace{0.5cm} &\eqref{Eqn:power_cons},~\eqref{Eqn:analog_cons}.
\end{align}
Here, the CHP subproblem is a non-convex problem with the constant-modulus constraints and coupled variables.
Note that, as a connection-state matrix ${\bf D}$ is assumed to be known in CHP subproblem, the hybrid precoders obtained from CHP subproblem \eqref{Eqn:problemCHP1} can be viewed as functions of connection-state matrix, denoted as $\{{\bf W}^\star({\bf D}), {\bf F}^\star({\bf D})\}$.
Thus, if we can efficiently solve CHP problem and  apply its solution $\{{\bf W}^\star({\bf D}), {\bf F}^\star({\bf D})\}$ into \eqref{Eqn:problemJSEE_maxpower}, the original problem  {can be equivalently} reformulated into a discrete connection-state (DCS)  {problem} shown as follows
\begin{align}\label{Eqn:problem_switch}
\max_{{\bf D}} \hspace{0.5cm} & {{\rm EE}}\big({\bf W}^\star({\bf D}), {\bf F}^\star({\bf D}),{\bf D}\big)\nonumber\\
\hspace{0cm} {\rm s.t.} \hspace{0.5cm} &\eqref{Eqn:power_cons}, \eqref{Eqn:switch_cons},\eqref{Eqn:validRFAnta_cons}, \eqref{Eqn:validantanna_RF_cons}.
\end{align}
Here, the DCS problem becomes a combinatorial optimization problem with a finite feasible region.


\textcolor{black}{ {Inspired by the above procedure}, the original problem \eqref{Eqn:problemJSEE_maxpower} can be solved with a two-step manner: 1)  {With a given connection-state matrix $\mathbf{D}$}, we derive the corresponding hybrid precoder by solving CHP subproblem \eqref{Eqn:problemCHP1}; 2)  {By treating the hybrid precoder obtained from CHP subproblem} \eqref{Eqn:problemCHP1}  {as a function of ${\bf D}$}, the connection-state matrix and the hybrid precoder can be jointly determined by solving DCS problem \eqref{Eqn:problem_switch}.}

\section{Joint Optimization of Hybrid Precoding and Switch-Controlled Connections}
\label{sec:algorithm}
 {In this section, we first propose an AHP algorithm to solve CHP subproblem  with a given connection matrix $\mathbf{D}$. Then, using the AHP algorithm to determine the utility of matching between the RF chains and antennas, an MA-FAHP algorithm is developed to solve the DCS problem.}

\subsection{AHP Algorithm for solving the CHP Subproblem}\label{Sec:CHP}

\textcolor{black}{Recall that CHP subproblem \eqref{Eqn:problemCHP1} is a non-convex problem with the constant-modulus and coupled variables. Thus, it is still challenging to solve this problem directly.
On the other hand, as pointed out by \cite{sparse_precoding,Zi2016JSAC}, the hybrid precoders can be optimized by minimizing Euclidean distances between itself and the optimal full-digital precoder.}

Targeting at the EE maximization, the full-digital precoding problem can be formulated as
\begin{align}\label{Eqn:problemFDP}
\max_{{\bf W}_{\rm FD}}\hspace{0.5cm} & \frac{{\rm log}_2\Big[{\rm det}\Big({\bf I}
+\frac{{\bf H}{\bf W}_{\rm FD}{\bf W}_{\rm FD}^H{\bf H}^H}{L_s\sigma^2}\Big)\Big]}{||{\bf W}_{\rm FD}||_{\rm F}^2/L_s+P_c({\bf D})}\nonumber\\
\hspace{0.2cm} {\rm s.t.} \hspace{0.5cm} &||{\bf W}_{\rm FD}||_{\rm F}^2\leq P_{\rm max}.
\end{align}
For the above precoding problem, the optimal full-digital precoder, denoted as ${\bf W}_{\rm FD}^{\rm opt}$, can be efficiently obtained by using Dinkelbach's transform based two-layer algorithm \cite{HeTCOM2013}. Then, with the optimal full-digital precoder, the optimal hybrid precoders can be obtained by minimizing the  Euclidean distance between the optimal full-digital and hybrid precoders, leading to the following problem
\begin{align}\label{Eqn:problemHybrid2}
\min_{{\bf W}, {\bf F}} \hspace{0.5cm}  & ||{\bf W}_{\rm FD}^{\rm opt}-({\bf F} \circ{\bf D}){\bf W}||_{\rm F}^2,\nonumber\\
\hspace{1.0cm} {\rm s.t.} \hspace{0.5cm} &\eqref{Eqn:power_cons},~\eqref{Eqn:analog_cons}.
\end{align}
However, due to coupled constraint \eqref{Eqn:power_cons} and non-convex constraint \eqref{Eqn:analog_cons}, it is still challenging to jointly optimize variables ${\bf F}$ and ${\bf W}$ in problem \eqref{Eqn:problemHybrid2}.
On the other hand, \textcolor{black}{by} removing coupled constraint \eqref{Eqn:power_cons}, \textcolor{black}{it} yields a relaxed problem given by
\begin{align}\label{Eqn:problemJSEE_DTminobj}
\min_{{\bf W}, {\bf F}} \hspace{0.2cm}  ||{\bf W}_{\rm FD}^{\rm opt}-({\bf F} \circ{\bf D}){\bf W}||_{\rm F}^2,
\hspace{3mm} {\rm s.t.} \hspace{0.2cm}\eqref{Eqn:analog_cons}.
\end{align}
For this relaxed problem, we have the following lemma to demonstrate its relation to problem \eqref{Eqn:problemHybrid2}.
\begin{lemma}\label{Lem:power_remove}
All KKT solutions of the relaxed problem \eqref{Eqn:problemJSEE_DTminobj} satisfy the power constraint \eqref{Eqn:power_cons}.
\end{lemma}
\begin{IEEEproof}
Following the proof of \cite[{Theorem 1}]{Jin2018TWC}, this lemma is obtained.
\end{IEEEproof}

As known from \textbf{\emph{Lemma}} \ref{Lem:power_remove}, the power constraints in problem \eqref{Eqn:problemHybrid2} can be removed, yielding a simplified problem as shown in (\ref{Eqn:problemJSEE_DTminobj}). In the following, we will propose an AHP algorithm to solve problem \eqref{Eqn:problemJSEE_DTminobj}.

The idea of the proposed AHP algorithm is to repeat the following two steps until convergence:  First, under a given digital precoder, the analog precoder can be optimized by using the existing numerical algorithms,  {such as the L-BFGS algorithm} \cite{LiuLBFGS2014}; Second, with the derived analog precoder, the  {closed-form optimal digital precoder can be obtained and used} as the given digital precoder for the next round. The detailed rationale of each step is described as follows.

\textbf{First step: }
When the digital precoder is given, problem \eqref{Eqn:problemJSEE_DTminobj} can be viewed as an analog precoding problem given by
\begin{align}\label{Eqn:problemJSEE_analog}
\min_{{\bf F}} \hspace{2mm} ||{\bf W}_{\rm FD}^{\rm opt}-({\bf F} \circ{\bf D}){\bf W}||_{\rm F}^2, \hspace{2mm} {\rm s.t.} \hspace{2mm}\eqref{Eqn:analog_cons}.
\end{align}
Then, using the property of the Kronecker Product that $({\bf B}^T\otimes{\bf A}){\rm{vec}}({\bf X})={\rm{vec}}(\bf{AXB})$, analog precoding problem \eqref{Eqn:problemJSEE_analog} can be reformulated as a vector form given by
\begin{subequations}\label{Eqn:problemAnalogVec}
\begin{align}
\min_{{\bf f}} \hspace{0.3cm}&||{\bf w}_{\rm FD}^{\rm opt}-{\bf A}_{Fa}({\bf f} \circ{\bf d})||_2^2,\\
\hspace{0cm} {\rm s.t.} \hspace{0.3cm}& |{\bf f}(i)|=\left\{
\begin{array}{ll}
1/\sqrt{M_{\rm T}}, & \textrm{if }{\bf {d}}(i)=1,
\\
0, & \textrm{if }{\bf {d}}(i)=0,
\end{array}
\right.
\label{eq:vec_constraint}
\end{align}
\end{subequations}
where ${\bf w}_{\rm FD}^{\rm opt}\triangleq{\rm{vec}}({\bf W}_{\rm FD}^{\rm opt})$, ${\bf f}\triangleq {\rm vec}({\bf F})$,  ${\bf d}\triangleq{\rm{vec}}({\bf D})$, and ${\bf A}_{Fa}={\bf W}^T\otimes {\bf I}_{M_{\rm T}}$.
As seen from constraint \eqref{eq:vec_constraint}, $|{\bf f}(i)|=0$ holds for ${\bf d}(i)=0$, meaning that ${\bf f}(i)$ is a zero element of vectorized analog precoder ${\bf f}$ when ${\bf d}(i)=0$ holds. Therefore, by removing all zero elements in vectorized analog precoder ${\bf f}$, vector-form analog precoding problem \eqref{Eqn:problemAnalogVec} can be further  {expressed} as
\begin{subequations}\label{Eqn:problemAnaValid}
\begin{align}
\min_{{\bf f}_v} \hspace{0.3cm}&||{\bf w}_{\rm FD}^{\rm opt}-{\bf A}_{fv}{\bf f}_v||_2^2,\\
\hspace{0cm} {\rm s.t.} \hspace{0.3cm} & |{\bf f}_v(i)|=\frac{1}{\sqrt{M_{\rm T}}},\hspace{0.3cm}{\rm{for}} ~~i=1,\cdots, n_v,
\label{eq:constant_mod}
\end{align}
\end{subequations}
where $n_v$ is the number of non-zero elements in ${\bf d}$, vector ${\bf f}_v\in \mathcal{C}^{n_v\times 1}$ is obtained by removing all zero elements in vectorized analog precoder ${\bf f}$, and matrix ${\bf A}_{fv}\in \mathcal{C}^{M_{\rm T}L_s \times n_v}$ is obtained by removing the $\tilde k$th column of the matrix ${\bf A}_{Fa}$ for all ${\tilde k}\in\mathcal{K}\triangleq \{{j|{\bf d}(j)=0}, j\in\{1,...,M_{\rm T}N_{\rm T}^{\rm RF}\}\}$.

Apparently, the non-zero elements of vectorized analog precoder can be obtained by solving problem \eqref{Eqn:problemAnaValid}.
However, due to the presence of the non-convex constant-modulus constraints \eqref{eq:constant_mod}, it is still difficult to solve problem \eqref{Eqn:problemAnaValid} directly.
To deal with these constraints, \textcolor{black}{let us} introduce a new variable ${\bf{\Phi}}_v$ with satisfying ${\bf f}_{v}=\frac{1}{\sqrt{{M}_{\rm T}}}{\rm{e}}^{j{\bf{\Phi}}_v}$. Then, \textcolor{black}{upon} replacing ${\bf f}_{v}$ by $\frac{1}{\sqrt{{M}_{\rm T}}}{\rm{e}}^{j{\bf{\Phi}}_v}$, problem \eqref{Eqn:problemAnaValid} can be rewritten as
\begin{align}\label{Eqn:problemAnaPhi}
\min_{{\bf{\Phi}}_v} \hspace{3mm}&||{\bf w}_{\rm FD}^{\rm opt}-\frac{1}{\sqrt{{M}_{\rm T}}}{\bf A}_{fv}{\rm{e}}^{j{\bf{\Phi}}_v}||_2^2.
\end{align}
Since the number of the antennas is very large in massive MIMO mmWave systems, problem \eqref{Eqn:problemAnaPhi} is a large-scale unconstrained non-convex problem, which can be efficiently solved by existing numerical algorithms, such as the L-BFGS algorithm \cite{LiuLBFGS2014}.

According to the indices of zero element in ${\bf d}$, the vectorized analog precoder can be obtained by inserting zero elements into the derived solution to problem \eqref{Eqn:problemAnaPhi}. Finally, the solution to \eqref{Eqn:problemJSEE_analog} can be obtained by rewriting the vectorized analog precoder into a matrix form.

\textbf{Second step:} With the analog precoder derived from the first step, the digital precoding problem can be written as
\begin{align}\label{Eqn:problemDigital}
\min_{{\bf W}} \hspace{0.5cm}  & ||{\bf W}_{\rm FD}^{\rm opt}-({\bf F} \circ{\bf D}){\bf W}||_{\rm F}^2.
\end{align}
Since the above digital precoding problem  is an unconstrained convex problem, its optimal solution can be readily obtained as
\begin{align}\label{Eqn:Digital_optima}
{\bf W}=[({\bf F}\circ{\bf D})^H({\bf F}\circ{\bf D})]^{-1}({\bf F}\circ{\bf D})^H{\bf W}_{\rm FD}^{\rm opt}.
\end{align}

Overall,  alternatingly solving problem \eqref{Eqn:problemAnaPhi} by using existing numerical algorithms and  problem \eqref{Eqn:problemDigital} by using the {closed-form} optimal solution in \eqref{Eqn:Digital_optima}, the solution of hybrid precoding problem \eqref{Eqn:problemJSEE_DTminobj} can be obtained when the solutions to both problems \eqref{Eqn:problemAnaPhi} and \eqref{Eqn:Digital_optima} converge. \textcolor{black}{The proposed AHP algorithm is summarized in Algorithm \ref{Alg:Alter_lbfgs},  {where  $\epsilon_{\rm in}$ is the accuracy parameter for controlling the accuracy of AHP algorithm.}}

\begin{algorithm}[!hpbt]
{\baselineskip 16pt
\caption{The Proposed AHP Algorithm}
\label{Alg:Alter_lbfgs}
\begin{algorithmic}[1]
\STATE \textbf{Initialize}: For a given connection-state matrix ${\bf D}$, optimize full-digital precoder ${\bf W}_{\rm FD}^{\rm opt}$ by solving \eqref{Eqn:problemFDP} using Dinkelbach's transform based two-layer algorithm \cite{HeTCOM2013}. Then, randomly initialize analog precoder ${\bf F}$ with satisfying constraint \eqref{Eqn:analog_cons};
\WHILE {$|{\text{Obj}}_{\bf w}^n-{\text{Obj}}_{\bf f}^n|\geq \epsilon_{\rm in}$}
\STATE With the given analog precoder, update ${\bf W}^n$ according to \eqref{Eqn:Digital_optima}, and calculate the objective function of \eqref{Eqn:problemJSEE_DTminobj} as ${\text{Obj}}_{\bf w}^n$;
\STATE  With the given digital precoder, update ${\bf F}^n$ according to \eqref{Eqn:problemJSEE_analog}-\eqref{Eqn:problemAnaPhi}, and solve problem \eqref{Eqn:problemAnaPhi} by applying existing numerical algorithms, such as L-BFGS algorithm \cite{LiuLBFGS2014}. Then, calculate the objective function of \eqref{Eqn:problemJSEE_DTminobj} as ${\text{Obj}}_{\bf f}^n$;
\ENDWHILE
\RETURN{${\bf W}$ and ${\bf F}$.}
\end{algorithmic}
}
\end{algorithm}


%
%
%
%
%
%
%
%

\subsection{Proposed MA-FAHP algorithm}\label{Sec:SS}

\textcolor{black}{When hybrid precoders are offered by the AHP algorithm, the computational complexity of using the exhaustive search to solve DCS problem \eqref{Eqn:problem_switch} is $\mathcal{O}(2^{M_{\rm T}N_{\rm T}^{\rm RF}})$.} However, as the number of antennas $M_{\rm T}$ is sufficiently large due to the deployment of  massive MIMO, such exponential complexity is unacceptably high for practical massive MIMO mmWave systems. On the other hand, DCS problem \eqref{Eqn:problem_switch} can be viewed as a \emph{many-to-many matching problem} between RF chains and antennas, which can be efficiently solved by using Matching Theory \cite{Bayat2016matching,Han2017matching}.

According to Matching Theory \cite{Bayat2016matching,Han2017matching}, a many-to-many matching between RF chains and antennas can be defined as follows.
\begin{Def}[Many-to-Many Matching]\label{Def:matchphi}
A many-to-many matching between RF chains and antennas, denoted as $\Psi$, represents a mapping from the set of ${S}_{RF}\cup {{S}_{A}}\cup \{0\}$ to the set of all subsets of ${S}_{RF}\cup {{S}_{A}}\cup \{0\}$, with satisfying the following conditions: 1) $\Psi(i) \subseteq  {S}_{RF}, \forall i\in{S}_{A}$, 2) $\Psi(j) \subseteq  {S}_{A},\forall j\in{S}_{RF}$, 3) $|\Psi(i)|\leq p_i, \forall i\in{S}_{A}$,  4) $|\Psi(j)|\leq q_j,\forall j\in {S}_{RF}$,  and 5) $j \in \Psi(i) \Leftrightarrow i \in \Psi(j), \forall i\in{S}_{A}, j\in{S}_{RF}$.
\end{Def}

Here, condition 1) means that each antenna is matched with a certain portion of RF chains, condition 2) means that each RF chain is matched with a certain portion of antennas, condition 3) limits the size of $\Psi(i)$ no larger than $p_i$, accounting for \eqref{Eqn:validRFAnta_cons}, condition 4) limits the size of $\Psi(j)$ no larger than and $q_j$, accounting for constraint \eqref{Eqn:validantanna_RF_cons}, and condition 5) means that when the $i$th antenna is  matched with the $j$th RF chain, the $j$th RF chain is also matched to the $i$th antenna.

Based on Definition \ref{Def:matchphi}, solving DCS problem \eqref{Eqn:problem_switch} can be equivalently viewed as seeking for the \emph{best} many-to-many matching that maximizes the utility function defined as
\begin{equation}
\label{eq:utility}
{U}(\Psi)\triangleq {\rm EE}\big[{\bf D}(\Psi), {\bf W}^\star\big({\bf D}(\Psi)\big), {\bf F}^\star\big({\bf D}(\Psi)\big)\big],
\end{equation}
where ${\bf D}(\Psi)$ represents the connection-state matrix corresponds to $\Psi$. Note that hybrid precoder  {$\big\{{\bf W}^\star\big({\bf D}(\Psi)\big), {\bf F}^\star\big({\bf D}(\Psi)\big)\big\}$ in} \eqref{eq:utility}  {can be obtained from the AHP algorithm.}

According to Matching Theory\cite{Bayat2016matching,Han2017matching}, finding out the best many-to-many matching requires that all RF chains and all antennas maintain a number of \emph{preference \textcolor{black}{utilities}}. More precisely, assuming that the current matching is $\Psi$, the preference  {utilities of the $i$th antenna can be defined as}
\begin{equation}
  \label{eq:pre_value_antenna}
  PU_{i,\Lambda}\triangleq U\left({\Psi}_{i,\Lambda}\right), \forall \Lambda\subset{S}_{RF}, |\Lambda|\leq p_i,
\end{equation}
where ${\Psi}_{i,\Lambda}$ represents a new matching where the $i$th antenna is matched to a portion of RF chains $\Lambda$ with cardinality no more than $p_i$, while the matching conditions of antennas other than the $i$th antenna remains the same with current matching $\Psi$. Likewise, the preference  {utility of the $j$th RF chain can be defined as}
\begin{equation}
  \label{eq:pre_value_RF}
  PU_{j,\Omega}\triangleq U\left(\Psi_{j,\Omega}\right), \forall \Omega\subset{S}_{A}, |\Omega|\leq q_j,
\end{equation}
where $\Psi_{j,\Omega}$ represents a new matching that the $j$th RF chain is matched to a portion of antennas $\Omega$ with cardinality no more than $q_j$, while the matching conditions of RF chains other than the $j$th RF chains remains the same with current matching $\Psi$.
\begin{remark}
\label{remark:externalities}
  The matching problem between antennas and RF chains belongs to many-to-many matching with \emph{externalities},
  where the preference \textcolor{black}{utilities} of each antenna/RF chain  not only depend on the RF chains/antennas that it is matched with, but also depend on the matching conditions of the other antennas/RF chains.
\end{remark}

 As known from Remark \ref{remark:externalities}, due to the externalities feature, the preference \textcolor{black}{utilities} of each antenna/RF chain \textcolor{black}{are} dynamically changing \textcolor{black}{subject to} current matching condition, implying that the conventional matching algorithms with fixed preference \textcolor{black}{utilities} may not be proper for the considered matching problem.

\textcolor{black}{Based on the intrinsic characteristics of the many-to-many matching with externalities}, we propose a novel matching assisted fully-adaptive hybrid precoding (MA-FAHP) algorithm to solve \textcolor{black}{problem \eqref{Eqn:problem_switch}}.
In prior to demonstrating the proposed MA-FAHP algorithm, we first provide some basic concepts of proposed MA-FAHP algorithm as follows.
\begin{Def}\label{Def:swap_matching}
\emph{(Swap Operation and Swap Matching)} Consider that $[i,j]$ and $[i^\prime,j^\prime]$ are two matched antenna-RF chain pairs in matching $\Psi$ with satisfying $i\neq i^\prime$, $j\in \Psi(i) \cap \{S_{ RF}\setminus\Psi(i^\prime)\}$, $j^\prime \in \Psi(i^\prime)\cap\{S_{ RF}\setminus\Psi(i)\}$.
The \emph{swap operation} over $([i,j],[i^\prime,j^\prime])$ means that the $i$th and $i^\prime$th antennas will exchange their currently matched the $j$th and $j^\prime$th RF chains whilst keeping the other matching conditions same as the current matching $\Psi$. After the swapping operation over $([i,j],[i^\prime,j^\prime])$, we have a new matching expressed as
\begin{align}
\Psi_{([i,j],[i^\prime,j^\prime])}^{\rm swap}\triangleq\big\{\Psi \backslash\{[i,j],[i^\prime,j^\prime]\}\big\}\cup \big\{[i,j^\prime],[i^\prime,j]\big\},
\end{align}
called \emph{swap matching} over $([i,j],[i^\prime,j^\prime])$.
\end{Def}

\begin{Def}
\emph{(Joining-In Operation and Joining-In Matching)}
Consider that the $i$th antenna is not matched with the $j$th RF chain in matching $\Psi$, i.e., $j\notin\Psi(i)$.
The \emph{joining-in operation} over $[i,j]$ means that the $i$th antenna will be matched with the $j$th RF chain while keeping other matching conditions same as matching $\Psi$. After the joining-in operation, we have a new matching expressed as
\begin{align}
\Psi_{[i,j]}^{\rm join}\triangleq \Psi \cup \{[i,j]\},
\end{align}
called \emph{joining-in matching} over $[i,j]$.
\end{Def}

\begin{Def}
\emph{(Leaving Operation and Leaving Matching)}
Consider that the $i$th antenna is matched with the $j$th RF chain in matching $\Psi$, i.e., $j\in\Psi(i)$.
The \emph{leaving operation} over $[i,j]$ means that the $i$th antenna will not be matched with the $j$th RF chain while keeping other matching conditions same as matching $\Psi$. After the leaving operation, we have a new matching expressed as
\begin{align}
\Psi_{[i,j]}^{\rm leave}\triangleq \Psi \cup \{[i,j]\},
\end{align}
called \emph{leaving matching} over $[i,j]$.
\end{Def}

\begin{Def}\label{Def:OPU}
  \emph{[Operation Preference Utility (OPU)]}
  Consider that the current matching between antennas and RF chains is $\Psi$. The OPU of swap operation over $([i,j],[i^\prime,j^\prime])$ is defined as
  \begin{equation}\label{Eqn:OPU_swap}
    {\rm OPU}_{([i,j],[i^\prime,j^\prime])}^{\rm swap}\triangleq \left\{U\left(\Psi_{([i,j],[i^\prime,j^\prime])}^{\rm swap}\right)-U(\Psi)\right\}^+,
  \end{equation}
  for $i\neq i^\prime$, $j\in \Psi(i) \cap \{S_{ RF}\setminus\Psi(i^\prime)\}$ and $j^\prime \in \Psi(i^\prime)\cap\{S_{ RF}\setminus\Psi(i)\}$.
  The OPU of joining-in operation over $[i,j]$ is defined as
  \begin{equation}\label{Eqn:OPU_joining}
    {\rm OPU}_{[i,j]}^{\rm join}=\left\{U(\Psi_{[i,j]}^{\rm join})-U(\Psi)\right\}^+, \text{ for } j\notin\Psi(i).
  \end{equation}
  The OPU of leaving operation over $[i,j](j\in\Psi(i))$ is defined as
  \begin{equation}\label{Eqn:OPU_leaving}
    {\rm OPU}_{[i,j]}^{\rm leave}=\left\{U(\Psi_{[i,j]}^{\rm leave})-U(\Psi)\right\}^+,  \text{ for } j\in\Psi(i).
  \end{equation}
\end{Def}
\begin{remark}
  \textcolor{black}{As known from \eqref{eq:utility}, the utility function is related to the solution of CHP problem \eqref{Eqn:problemCHP1}, which can be obtained from the proposed AHP algorithm. Thus, computing the OPUs defined above needs to use the proposed AHP algorithm with connection-state matrices that corresponds to swap/joining-in/leaving matching and the current matching $\Psi$.}
\end{remark}


With above Definitions \ref{Def:swap_matching}$\sim$\ref{Def:OPU}, the proposed MA-FAHP algorithm is summarized in Algorithm \ref{Alg:swap-matching}. The main idea of the proposed MA-FAHP algorithm is demonstrated as follows.
In each ``while'' loop, 
by computing the OPUs defined in (\ref{Eqn:OPU_swap}), (\ref{Eqn:OPU_joining}) and (\ref{Eqn:OPU_leaving}),  the \emph{best} swap operation, the \emph{best} joining-in operation and the \emph{best} leaving operation  are  sequentially found out and then executed for every antenna. As a result, the many-to-many matching is optimized in each ``while'' loop in order to achieve a higher utility value than the previous one. Finally, a stable matching that can not be further improved by any swap/joining-in/leaving operation will be achieved after several ``while'' loops, which will be theoretically proved in Section \ref{Sec:perforAnaly}.
\begin{algorithm}[!t]
{\baselineskip 16pt
\caption{The Proposed MA-FAHP Algorithm}
\label{Alg:swap-matching}
\begin{algorithmic}[1]
\STATE\textbf{Initialization}: Initialize the many-to-many matching $\Psi$ based on a randomly generated connection-state matrix ${\bf D}$; 
 Let the stability indicator be ${ \delta}_{\rm stab}=0$;
\WHILE{ (${ \delta}_{\rm stab}=0$) }
\STATE $\Psi_{\rm pre}\leftarrow \Psi$;
  \FOR {$i = 1$ to $M_{\rm T}$}
  \STATE
  According to \eqref{Eqn:OPU_swap}, compute the OPUs of all possible swap operations involving the $i$th antenna by using the AHP algorithm.
  Find out the best swap operation that achieves the highest OPU by
  \begin{equation}
  \label{eq:best_swap}
    ([i,j],[i^\prime,j^\prime])^\star=
    \arg\max_{\substack {i^\prime\neq i, i^\prime\in S_{A}\\j\in\Psi(i)\cap \{S_{ RF}\setminus\Psi(i^\prime)\},\\j^\prime\in\Psi(i^\prime) \cap\{S_{ RF}\setminus\Psi(i)\}}} {\rm OPU}_{([i,j],[i^\prime,j^\prime])}^{\rm swap}
  \end{equation}
  If ${\rm OPU}_{ ([i,j],[i^\prime,j^\prime])^\star}^{\rm swap}>0$, execute the swap operation over $([i,j],[i^\prime,j^\prime])^\star$ and update $\Psi\leftarrow\Psi_{ ([i,j],[i^\prime,j^\prime])^\star}^{\rm swap}$;

\STATE 
According to \eqref{Eqn:OPU_joining}, compute OPUs of all possible joining-in operations involving the $i$th antenna by using the AHP algorithm.
  Find out the best joining-in operation that achieves the highest OPU by
  \begin{equation}
  \label{eq:best_join}
    [i,j]^\star=\arg\max_{j\notin\Psi(i)}{\rm OPU}_{[i,j]}^{\rm join}
  \end{equation}
  If ${\rm OPU}_{ [i,j]^\star}^{\rm join}>0$, execute the joining-in operation over $[i,j]^\star$ and update $\Psi\leftarrow\Psi_{[i,j]^\star}^{\rm join}$;

\STATE 
According to \eqref{Eqn:OPU_leaving}, compute OPUs of all possible leaving operations involving the $i$th antenna by using the AHP algorithm .
  Find out the best leaving operation that achieves the highest OPU by
  \begin{equation}
  \label{eq:best_leave}
    [i,j]^\star=\arg\max_{j\in\Psi(i)}{\rm OPU}_{[i,j]}^{\rm leave}
  \end{equation}
  If ${\rm OPU}_{ [i,j]^\star}^{\rm leave}>0$, execute the leaving operation over $[i,j]^\star$ and update $\Psi\leftarrow\Psi_{[i,j]^\star}^{\rm leave}$;

  \ENDFOR
\IF {($\Psi==\Psi^{\rm pre}$)}
\STATE $\delta_{\rm stab}\leftarrow 1$;
\ENDIF
\ENDWHILE
\RETURN \textcolor{black}{Generate connection-state matrix ${\bf D}$ according to $\Psi$; Then, apply the generated ${\bf D}$ into AHP algorithm to obtain the hybrid precoders $\{{\bf W}^\star ({\bf D}), {\bf F}^\star ({\bf D})\}$.}
\end{algorithmic}
}
\end{algorithm}

\section{Convergence and Complexity}\label{Sec:perforAnaly}


This section theoretically analyzes the convergence and complexity of the proposed MA-FAHP algorithm.

\subsection{Convergence}

\begin{Prop}\label{Pro:convergence_global}
The proposed MA-FAHP algorithm 
 converges to a stable matching.
\end{Prop}

\begin{IEEEproof}
Recall that when the OPU of the best swap/joining-in/leaving operation is a positive constant, the corresponding operation will be executed in each ``while'' loop. Therefore, the utility of matching keeps monotonically increasing loop by loop.
On the other hand, recall that the $i$th antenna can be matched with up to a number $p_i$ of RF chains, while the $j$th RF chain can be matched with up to a number $q_j$ of antennas. Thus, the sample space of matchings is finite, indicating that the utility values of all possible conditions of many-to-many matching are upper bounded.
From the non-decreasing monotonicity and upper boundedness demonstrated above, it can be concluded that the utility value monotonically increases and eventually converges to a stable matching,  {whose utilities of the best swap operation, the best joining-in operation and the best leaving operation for all antennas are all equal to zero.} This completes the proof.
\end{IEEEproof}

\subsection{Complexity}
\label{sec:complexity}

Since L-BFGS algorithm can efficiently solve the unconstrained large-scale non-convex  problem, the proposed AHP algorithm is expected to efficiently solve CHP problem \eqref{Eqn:problem_switch} with reasonable complexity. Therefore, in the following, we do not consider the complexity of using AHP algorithm to determine hybrid analog/digital precoders.

In each ``while'' loop of proposed MA-FAHP algorithm, the computational complexity mainly comes from finding out the best swap operation, the best joining-in operation and the best leaving operation.
As seen from \eqref{eq:best_swap}, the complexity for the $i$th antenna to find out the best swap operation can be expressed as
\begin{equation}
\sum_{i^\prime\neq i}|\Psi(i)\cap \{S_{ RF}\setminus\Psi(i^\prime)\}|\times|\Psi(i^\prime)\cap \{S_{ RF}\setminus\Psi(i)\}|
\overset{\rm (i)}{\leq}
\sum_{i^\prime\neq i} |\Psi(i)|\times |\Psi(i^\prime)|\overset{\rm (ii)}{\leq}  \sum_{i^\prime\neq i} p_i p_{i^\prime},
\end{equation}
where step (i) uses the fact that $|\Psi(i)\cap \{S_{ RF}\setminus\Psi(i^\prime)\}|\leq|\Psi(i)|$ and
$|\Psi(i^\prime)\cap \{S_{ RF}\setminus\Psi(i)\}|\leq |\Psi(i^\prime)|$, and step (ii) comes from $|\Psi(i)|\leq p_i$ and $|\Psi(i^\prime)|\leq p_{i^\prime}$.
Further,  it can be observed from \eqref{eq:best_join} that the complexity of finding out the best joining-in operation for the $i$th antenna is $|\Psi(i)|$. Likewise, the complexity incurred by finding out the best leaving operation for the $i$th antenna is $N_{\rm T}^{\rm RF}-|\Psi(i)|$, as observed from \eqref{eq:best_leave}.
Recall that in each ``while'' loop of MA-FAHP algorithm, the best swap operation, the best joining-in operation  and the best leaving operation should be found out for \emph{every} antenna. Therefore, the computational complexity of each ``while'' loop is upper bounded by
\begin{multline}
  \sum_{i=1}^{M_{\rm T}} \left[\sum_{i^\prime\neq i} p_i p_{i^\prime}+ |\Psi(i)|+\left(N_{\rm T}^{\rm RF}-|\Psi(i)|\right)\right]
  \\
  = \left(\sum_{i=1}^{M_{\rm T}} \sum_{i^\prime\neq i} p_i p_{i^\prime}\right)+ M_{\rm T} N_{\rm T}^{\rm RF}\overset{\rm (iii)}{=}\mathcal{O}\left[M_{\rm T}^{2} (N_{\rm T}^{\rm RF})^2\right]
\end{multline}
where step (iii) uses the fact $p_i\leq N_{\rm T}^{\rm RF}$ holds for $i=1,...,M_{\rm T}$.
Supposing that the MA-FAHP algorithm reaches a stable matching after $\Delta$ ``while'' loops,  the overall computational complexity \textcolor{black}{required by the} MA-FAHP algorithm is up to $\mathcal{O}\left[\Delta M_{\rm T}^{2} (N_{\rm T}^{\rm RF})^2\right]$, which is much less than the exponential complexity $\mathcal{O}(2^{M_{\rm T} N_{\rm T}^{\rm RF}})$ incurred by the exhaustive search.

\section{Numerical Simulation}\label{Sec:simulation}

This section presents numerical results to demonstrate the performance of proposed hybrid precoding with a fully-adaptive-connected structure. In our simulation, the power consumptions of each working RF chain, each working phase shifter, each switch and other circuit components are set to  $P_{\rm RF}=0.3$W, $P_{\rm PS}=0.05$W, $P_{\rm SW}=0.01$W, and $P_{o}=0.5$W, respectively. Without loss of generality, the channel between the source and the user experiences normalized Rayleigh fading with unit variance, and the noise power is set to ${\sigma}^2=1$W. Moreover, the accuracy parameter of AHP algorithm is set to $\epsilon_{\rm in}=10^{-4}$.
The simulation results are averaged from 1000 independently random channel realizations.

In our simulation, the fully-adaptive-connected precoding using MA-FAHP algorithm, termed as \textcolor{black}{\emph{proposed precoding strategy}} hereafter, is compared with four existing precoding strategies listed as follows
\begin{itemize}
\item Strategy-1: full-digital precoding that uses the Dinkelbach's transform based two-layer algorithm for EE maximization \cite{HeTCOM2013};
\item Strategy-2: full-connected precoding that uses the energy efficient hybrid precoding (EEHP) algorithm for EE maximization \cite{Zi2016JSAC};
\item Strategy-3: sub-connected precoding that uses the semidefinite relaxation-alternating minimization (SDR-AltMin) algorithm \cite{Alter_Hybrid} to solve problem \eqref{Eqn:problemHybrid2} under a given connection-state matrix\footnote{Following the setup in \cite{Alter_Hybrid}, the connection-state matrix is set to represent the case that all antennas are equally partitioned into $N_{\rm T}^{\rm RF}$ groups and the $j$th RF chain is connected to the $j$th antenna group, where $j=1,...,N_{\rm T}^{\rm RF}$.}.
\item Strategy-4: fully-adaptive-connected precoding that uses the phase shifter selection scheme \cite{Payami2016TWC} for SE maximization\footnote{Although Strategy-4 targets at the SE maximization, we still compare our proposed hybrid precoding strategy with Strategy-4, as the both employs a fully-adaptive-connected structure.}.
\end{itemize}

\begin{figure}
\centering
\includegraphics[width=0.5\textwidth]{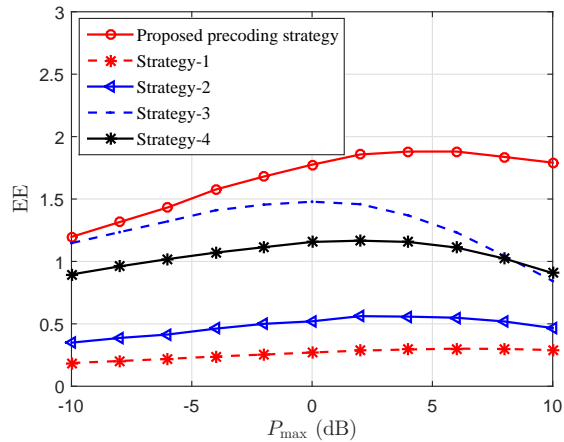}
\caption{The EE comparison among the proposed  precoding strategy and Strategies 1$\sim$4 versus the maximal transmit power $P_{\rm max}$, where $M_{\rm T}=M_{\rm R}=64$, $N_{\rm T}^{\rm RF}=4$, and $L_s=1$.}
\label{Fig:JSEE}
\end{figure}
Fig. \ref{Fig:JSEE} depicts the EE of proposed  precoding strategy and Strategies-1$\sim$4 versus the maximal transmit power $P_{\rm max}$. As seen from this figure,  the proposed precoding strategy achieves higher EE than Strategies-1$\sim$4. The reason for this observation can be explained as follows.
First, as the full-digital precoding and full-connected hybrid precoding structures employ much more RF chains and phase shifters than the fully-adaptive-connected structure,
Strategies-1$\sim$2 incur much higher circuit power consumption than the proposed precoding strategy. Second, since each switch consumes much less than each phase shifter, the proposed precoding strategy also consumes less circuit power than Strategy-3 that employs the sub-connected structure.
Moreover, since the phase shifter selection scheme used in Strategy-4 aims at maximizing the SE without considering the power consumption, Strategy-4 also achieves lower EE than the proposed precoding strategy.

\begin{figure}
\centering
\includegraphics[width=0.5\textwidth]{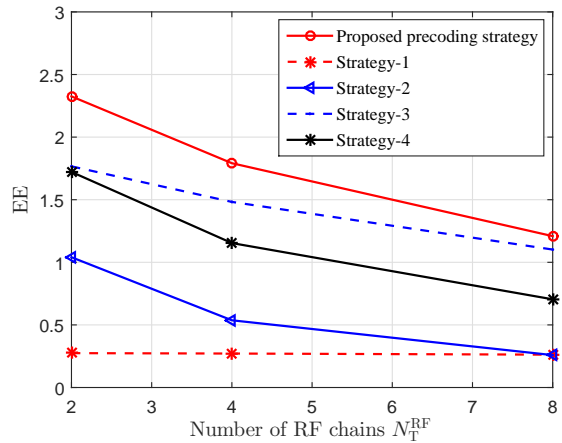}
\caption{The EE comparison among the proposed  precoding strategy and Strategies 1$\sim$4 versus the number of RF chains $N_{\rm T}^{\rm RF}$, where $M_{\rm T}=M_{\rm R}=64$, $P_{\rm max}=10$ dBW, and $L_s=1$.}
\label{Fig:rf_chains}
\end{figure}
Fig. \ref{Fig:rf_chains} demonstrates the EE of proposed  precoding strategy and Strategies-1$\sim$4 versus the number of RF chains $N_{\rm T}^{\rm RF}$.
It can be seen that the proposed  precoding strategy outperforms all other precoding schemes in terms of EE. Further, it can also be seen that, as the number of RF chain increases, the EE gap between the proposed precoding strategy and Strategy-3 becomes smaller. This is because when $N_{\rm T}^{\rm RF}$ is sufficiently large, the power consumption in \eqref{Eqn:powerconsumptionBS_final} is dominated by the RF chains, indicating that the fully-adaptive-connected structure and the sub-connected structure consumes similar circuit power.


\begin{figure}
\centering
\includegraphics[width=0.5\textwidth]{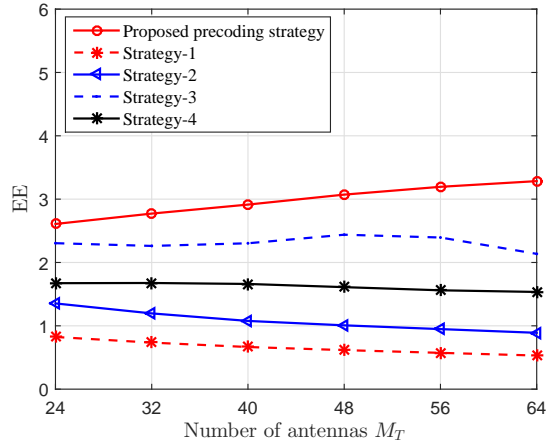}
\caption{The EE comparison among the proposed  precoding strategy and Strategies 1$\sim$4 versus the number of antennas $M_{\rm T}$, where $N_{\rm T}^{\rm RF}=4$, $P_{\rm max}=10$ dBW, and $L_s=2$.}
\label{Fig:antennas}
\end{figure}
Fig. \ref{Fig:antennas} demonstrates the EEs of proposed precoding strategy and Strategies-1$\sim$4 versus the number of antennas $M_{\rm T}$.
As observed from this figure, the proposed precoding strategy outperforms Strategies-1$\sim$4 in terms of EE. Moreover, when the number of antennas further increases, the proposed precoding strategy achieves increasing EE, whereas Strategies-1$\sim$2 and Strategy-4 achieve decreasing EE, and Strategy-3 achieves fluctuating EE.
This observation indicates that the fully-adaptive-connected structure is more suitable for massive MIMO systems which employs a large number of antennas.

\begin{figure}
\centering
\includegraphics[width=0.5\textwidth]{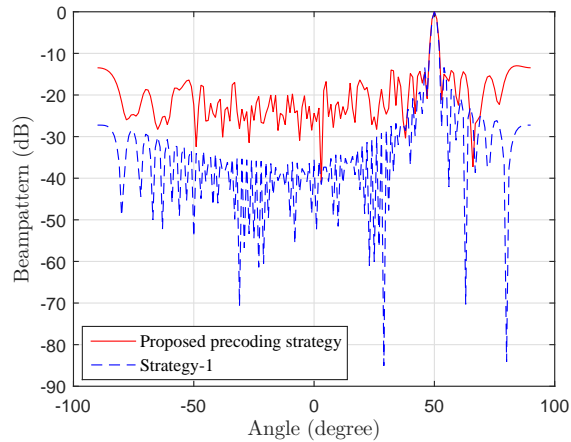}
\caption{The beampatterns in the cartesian coordinate system for the proposed hybrid precoding and the full-digital precoding, where  $N_{\rm T}^{\rm RF}=2$, $M_{\rm T}=M_{\rm R}=64$ and $L_s=1$.}
\label{Fig:beam}
\end{figure}

\begin{figure}
\centering
\includegraphics[width=0.5\textwidth]{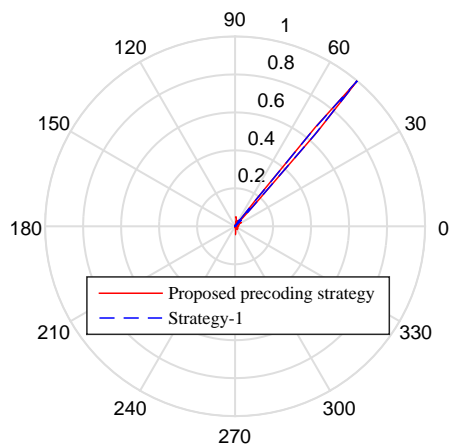}
\caption{The beampatterns in the polar coordinate system for the proposed hybrid precoding and the full-digital precoding, where $N_{\rm T}^{\rm RF}=2$, $M_{\rm T}=M_{\rm R}=64$ and $L_s=1$.}
\label{Fig:beam_polar}
\end{figure}

Figs. \ref{Fig:beam} and \ref{Fig:beam_polar} show the beampatterns of the proposed precoding strategy and Strategy-1 in Cartesian and polar coordinates, respectively, where the desired angle of departure (AoD) is $50^{\circ}$.
Here, the reason for only Strategy-1 being used for the comparison is that Strategy-1 is expected to provide the best beampattern among Strategies-1$\sim$4 due to its full-digital precoding structure.
It can be seen that both the proposed precoding strategy and Strategy-1 generate beams with direction at $50^{\circ}$, which complies with the desired AoD.
Further, as shown in Fig. \ref{Fig:beam}, the proposed precoding strategy has around 15 dB higher sidelobe than Strategy-1.
The reason for this observation is that the hybrid precoding structure employed in the proposed precoding strategy, has more sidelobe energy leakage than the full-digital structure employed in Strategy-1.
Nevertheless, the main lobe of the proposed precoding strategy achieves around 15 dB gain over its sidelobes, meaning that the proposed precoding  strategy still owns great directionality.

\begin{figure}
\centering
\includegraphics[width=0.5\textwidth]{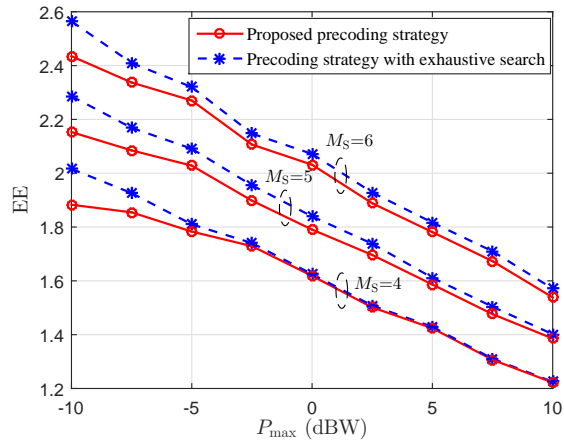}
\caption{\textcolor{black}{The EE comparisons of the proposed MA-FAHP algorithm and exhaustive search with using AHP algorithm, where  $N_{\rm T}^{\rm RF}=2$ and $L_s=1$.}}
\label{Fig:exhaustive}
\end{figure}

Finally, we evaluate the optimality of proposed MA-FAHP algorithm by comparing it with the exhaustive search.
Fig. \ref{Fig:exhaustive} compares the the proposed MA-FAHP algorithm and exhaustive search for solving the DCS problem \eqref{Eqn:problem_switch}. For making a fair comparison, the AHP algorithm is also used in exhaustive search for determining the hybrid precoder $\{{\bf W}^\star ({\bf D}), {\bf F}^\star ({\bf D})\}$, which is same as the proposed MA-FAHP algorithm. It can be observed from Fig. \ref{Fig:exhaustive} that when $M_{\rm T}=4$, the proposed MA-FAHP algorithm achieves almost the same EE with the exhaustive search when $P_{\rm max}\geq -2.5$ dBW. Further, as the the number of antennas increases from $M_{\rm T}=4$ to $M_{\rm T}=6$, the EE gap between the proposed MA-FAHP algorithm slightly increases to 0.04 with $P_{\rm max}\geq 0$ dBW, indicating that  the proposed MA-FAHP algorithm can still achieve the similar EE performance with the exhaustive search. Recall that the  MA-FAHP algorithm incurs polynomial complexity, which is much less than the exponential complexity of the exhaustive search, as demonstrated in Section \ref{sec:complexity}. Thus, it can be concluded that the MA-FAHP algorithm can achieve the similar EE with the exhaustive search while requiring much less complexity.


\section{Conclusion}
\label{sec:conclusion}
In this paper, we have studied massive MIMO mmWave hybrid precoding with a fully-adaptive-connected structure.
In order to maximize the EE, the joint optimization of hybrid precoding and connection-state matrix has been formulated as a large-scale high-dimensional mixed-integer non-convex problem with both continuous and discrete variables. To efficiently solve such a challenging problem, an MA-FAHP algorithm has been proposed to jointly determine the digital and analog precoders as well as the connection-state matrix. The convergence and the complexity of proposed MA-FAHP algorithm have been theoretically evaluated. Simulation results have demonstrated that the fully-adaptive-connected hybrid precoding using the proposed MA-FAHP algorithm can achieve better EE than the existing full-digital and hybrid precoding strategies.


\begin{thebibliography}{1}
\bibitem{XuanICC2019}
X. Xue, Y. Wang, L. Yang, J. Shi, and Zan Li, ``Spectral-energy
efficient hybrid precoding for mmWave systems with an adaptive-connected structure,'' in \emph{Proc. IEEE ICC}, May. 2019, pp. 1--6.
\bibitem{mmWaveCOMag2014Feb}
W. Roh, J-Y. Seol, J. Park, B. Lee, J. Lee, Y. Kim, J. Cho, K. Cheun, and F. Aryanfar, ``Millimeter-wave beamforming as an enabling technology for 5G cellular communications: Theoretical feasibility and prototype results,'' \emph{IEEE Commun. Mag.}, vol. 52, no. 2, pp. 106--113, Feb. 2014.
\bibitem{mmwave_introduction}
M. Xiao, S. Mumtaz, Y. Huang, L. Dai, Y. Li, M. Matthaiou, G. K. Karagiannidis, E. Bj\"{o}rnson, K. Yang, C. L. I, and A. Ghosh, ``Millimeter wave communications for future mobile networks,'' \emph{IEEE J. Sel. Areas Commun.}, vol.~35, no.~9, pp.~1909--1935, Sep. 2017.
\bibitem{mm-wave_it_works}
T. S. Rappaport, S. Sun and R. Mayzus, H. Zhao, Y. Azar, K. Wang, G. N. Wong, J. K. Schulz, M. Samimi and F. Gutierrez, ``Millimeter wave mobile communications for 5G cellular: It will work!'', \emph{IEEE Access}, vol.~1, pp.~335--349, May 2013.
\bibitem{HeathJSTSP2016}
R. W. Heath, N. Gonz\'{a}lez-Prelcic, S. Rangan, W. Roh and A. M. Sayeed, ``An overview of signal processing techniques for millimeter wave MIMO systems,'' \emph{IEEE J. Sel. Topics Signal Process.}, vol. 10, no. 3, pp. 436--453, Apr. 2016.
\bibitem{HanComMag2015}
S. Han, C. I, Z. Xu and C. Rowell, ``Large-scale antenna systems with hybrid analog and digital beamforming for millimeter wave 5G,'' \emph{IEEE Commun. Mag.}, vol. 53, no. 1, pp. 186--194, Jan. 2015.
\bibitem{mm-wave_potential}
S. Mumtaz, J. Rodriquez, and L. Dai, \emph{MmWave Massive MIMO: A Paradigm for 5G}, Academic Press., Elsevier, 2016.
\bibitem{AlkhateebComMag2014}
A. Alkhateeb, J. Mo, N. Gonzalez-Prelcic and R. W. Heath, ``MIMO precoding and combining solutions for millimeter-wave systems,'' \emph{IEEE Commun. Mag.}, vol. 52, no. 12, pp. 122--131, December 2014.
\bibitem{YooJSAC2006}
T. Yoo and A. Goldsmith, ``On the optimality of multiantenna broadcast scheduling using zero-forcing beamforming,'' \emph{IEEE J. Sel. Areas Commun.}, vol. 24, no. 3, pp. 528--541, Mar. 2006.
\bibitem{RanganProceed2014}
S. Rangan, T. S. Rappaport and E. Erkip, ``Millimeter-Wave Cellular Wireless Networks: Potentials and Challenges,'' \emph{Proceed. IEEE}, vol. 102, no. 3, pp. 366--385, Mar. 2014.
\bibitem{Hong2014ComMag}
W. Hong, K. Baek, Y. Lee, Y. Kim and S. Ko, ``Study and prototyping of practically large-scale mmWave antenna systems for 5G cellular devices,'' \emph{IEEE Commun. Mag.}, vol. 52, no. 9, pp. 63--69, Sep. 2014.

\bibitem{sparse_precoding}
O. E. Ayach, S. Rajagopal, S. Abu-Surra, Z. Pi, and R. W. Heath, ``Spatially sparse precoding in millimeter wave MIMO systems,'' \emph{IEEE Trans. Wireless Commun.}, vol.~13, no.~3, pp.~1499--1513, Mar. 2014.
\bibitem{Alter_Hybrid}
X. Yu, J. C. Shen, J. Zhang, and K. B. Letaief, ``Alternating minimization algorithms for hybrid precoding in millimeter wave MIMO systems," \emph{IEEE J. Sel. Topics Signal Process.}, vol.~10, no.~3, pp.~485--500, Apr. 2016.
\bibitem{chen2015WCL}
C. E. Chen, ``An iterative hybrid transceiver design algorithm for millimeter wave MIMO systems,'' \emph{IEEE Wireless Commun. Lett.}, vol. 4, no. 3, pp. 285--288, Jun. 2015.
\bibitem{Jin2018TWC}
J. Jin, Y. R. Zheng, W. Chen and C. Xiao, ``Hybrid precoding for millimeter wave MIMO systems: A matrix factorization approach,'' \emph{IEEE Trans. Wireless Commun.} vol.~17, no.~5, pp.~3327--3339, May 2018.
\bibitem{Alkhateeb2015TWC}
A. Alkhateeb, G. Leus and R. W. Heath, ``Limited feedback hybrid precoding for multi-user millimeter wave systems,'' \emph{IEEE Trans. Wireless Commun.}, vol. 14, no. 11, pp. 6481--6494, Nov. 2015.

\bibitem{Ni2016TCOM}
 W. Ni and X. Dong, ``Hybrid block diagonalization for massive multiuser MIMO systems,'' \emph{IEEE Trans. Commun.}, vol. 64, no. 1, pp. 201--211, Jan. 2016.

\bibitem{Liang2014WCL}
L. Liang, W. Xu and X. Dong, ``Low-complexity hybrid precoding in massive multiuser MIMO systems,'' \emph{IEEE Wireless Commun. Lett.}, vol. 3, no. 6, pp. 653--656, Dec. 2014.

\bibitem{Dai2015ICC}
L. Dai, X. Gao, J. Quan, S. Han, and C.-L. I, ``Near-optimal hybrid analog and digital precoding for downlink mmWave massive MIMO systems,'' in \emph{Proc. IEEE ICC}, Jun. 2015, pp. 1334--1339.

\bibitem{Gao2016JSAC}
X. Gao, L. Dai, S. Han, C.-L. I and R. W. Heath, ``Energy-efficient hybrid analog and digital precoding for mmWave MIMO systems with large antenna arrays,'' \emph{IEEE J. Sel. Areas Commun.}, vol. 34, no. 4, pp. 998-1009, Apr. 2016.
\bibitem{Du2018TWC}
J.~Du, W. Xu, H. Shen, X. Dong and C. Zhao, ``Hybrid precoding architecture for massive multiuser MIMO with dissipation: Sub-connected or fully connected structures?''  \emph{IEEE Trans. Wireless Commun.}, vol. 17, no. 8, pp.~5465--5479, Aug. 2018.

\bibitem{Yu2018JSTSP}
X. Yu, J. Zhang and K. B. Letaief, ``A hardware-efficient analog network structure for hybrid precoding in millimeter wave systems,'' \emph{IEEE J. Sel. Topics Signal Process.}, vol.~12, no.~2, pp. 282--297, May 2018.
\bibitem{Park2017TWC}
S. Park, A. Alkhateeb and R. W. Heath, ``Dynamic subarrays for hybrid precoding in wideband mmWave MIMO systems,''  \emph{IEEE Trans. Wireless Commun.}, vol. 16, no. 5, pp. 2907--2920, May 2017.
\bibitem{Jing2018ComL}
X. Jing, L. Li, H. Liu and S. Li, ``Dynamically connected hybrid precoding scheme for millimeter-wave massive MIMO systems,'' \emph{IEEE Commun. Lett.}, vol. 22, no. 12, pp. 2583--2586, Dec. 2018.
\bibitem{Payami2016TWC}
S. Payami, M. Ghoraishi and M. Dianati, ``Hybrid beamforming for large antenna arrays with phase shifter selection,'' \emph{IEEE Trans. Wireless Commun.}, vol. 15, no. 11, pp. 7258--7271, Nov. 2016.
\bibitem{Yazdan2017MicMag}
A. Yazdan, J. Park, S. Park, T. A. Khan and R. W. Heath, ``Energy-efficient massive MIMO: Wireless-powered communication, multiuser MIMO with hybrid precoding, and cloud radio access network with variable-resolution ADCs,'' \emph{IEEE Microwave Mag.}, vol. 18, no. 5, pp. 18--30, July-Aug. 2017.
\bibitem{He2017TSP}
S. He, J. Wang, Y. Huang, B. Ottersten and W. Hong, ``Codebook-based hybrid precoding for millimeter wave multiuser systems,'' \emph{IEEE Trans. on Signal Process.}, vol. 65, no. 20, pp. 5289--5304, Oct. 2017.
\bibitem{He2016access}
S. He, C. Qi, Y. Wu and Y. Huang, ``Energy-efficient transceiver design for hybrid sub-array architecture MIMO systems," \emph{IEEE Access}, vol. 4, pp. 9895--9905, 2016.
\bibitem{Zi2016JSAC}
R. Zi, X. Ge, J. Thompson, C. Wang, H. Wang and T. Han, ``Energy efficiency optimization of 5G radio frequency chain systems,'' \emph{IEEE J. Sel. Areas Commun.}, vol. 34, no. 4, pp. 758--771, Apr. 2016.

\bibitem{NgoTCOM2013}
H. Q. Ngo, E. G. Larsson and T. L. Marzetta, ``Energy and spectral efficiency of very large multiuser MIMO systems,'' \emph{IEEE Trans. Commun.}, vol. 61, no. 4, pp. 1436--1449, Apr. 2013.

\bibitem{HeTCOM2013}
S. He, Y. Huang, S. Jin and L. Yang, ``Coordinated beamforming for energy efficient transmission in multicell multiuser Systems,'' \emph{IEEE Trans. Commun.}, vol. 61, no. 12, pp. 4961--4971, Dec. 2013.

\bibitem{LiuLBFGS2014}
T. W. Liu. ``A regularized limited memory BFGS method for nonconvex unconstrained minimization,'' \emph{Numerical Algorithms}, vol.~65, no.~2, pp.~305--323, Oct. 2014.

\bibitem{Bayat2016matching}
S. Bayat, Y. Li, L. Song and Z. Han, ``Matching theory: Applications in wireless communications,'' \emph{IEEE Signal Process. Mag.}, vol. 33, no. 6, pp. 103--122, Nov. 2016.

\bibitem{Han2017matching}
Z. Han, Y. Gu, and W. Saad, \emph{Matching Theory for Wireless Networks}, Incorporated Springer Publishing Company, 2017.

\bibitem{matching_sub_channel_assignment}
B. Di, L. Song and Y. Li, ``Sub-Channel assignment, power allocation, and user scheduling for non-orthogonal multiple access networks,'' \emph{IEEE Trans. Wireless Commun.}, vol. 15, no. 11, pp. 7686--7698, Nov. 2016.

\bibitem{matching_many_one}
E. Bodine-Baron, C. Lee, A. Chong, B. Hassibi, and A. Wierman, ``Peer effects and stability in matching markets,'' Proc. 4th Symp. Algorithmic Game Theory (SAGT), Amalfi, Italy, Oct. 2011, pp. 117--129.

\bibitem{mmwav_wireless}
T. S. Rappaport, R. W. Heath, R. C. Daniels, and J. N. Murdock, \emph{Millimeter wave wireless communications}, Prentice Hall, 2015.
\end{thebibliography}
\end{document}